\documentclass[aps,prl,onecolumn,amsmath,amssymb,superscriptaddress]{revtex4}
\usepackage{graphicx}
\usepackage{graphics}
\usepackage{bm}
\usepackage[usenames]{color}
\usepackage{colordvi}
\usepackage{units}
\usepackage{bbm}
\usepackage{booktabs}
\usepackage{array}
\usepackage{placeins}
\usepackage{epsfig}
\usepackage{lscape}
\usepackage{changes}
\usepackage{braket}
\usepackage{float}
\usepackage{diagbox}
\usepackage{layouts}

\bibstyle{apsrev.bib}

\setcitestyle{round}

\begin{document}

\title{Spatial cytoskeleton organization supports targeted intracellular transport}

\author{Anne E. Hafner}
\author{Heiko Rieger}
\email{h.rieger@mx.uni-saarland.de}
\affiliation{Department of Theoretical Physics, Saarland University, 66123 Saarbr\"ucken, Germany}



\begin{abstract}
\noindent
The efficiency of intracellular cargo transport from specific source to target locations is strongly dependent upon molecular motor-assisted motion along the cytoskeleton. Radial transport along microtubules and lateral transport along the filaments of the actin cortex underneath the cell membrane are characteristic for cells with a centrosome. The interplay between the specific cytoskeleton organization and the motor performance realizes a spatially inhomogeneous intermittent search strategy. In order to analyze the efficiency of such intracellular search strategies we formulate a random velocity model with intermittent arrest states. We evaluate efficiency in terms of mean first passage times for three different, frequently encountered intracellular transport tasks: 
i) the narrow escape problem, which emerges during cargo transport to a synapse or other specific region of the cell membrane, 
ii) the reaction problem, which considers the binding time of two particles within the cell, and
iii) the reaction-escape problem, which arises when cargo must be released at a synapse only after pairing with another particle. 
Our results indicate that cells are able to realize efficient search strategies for various intracellular transport tasks economically through a spatial cytoskeleton organization that involves only a narrow actin cortex rather than a cell body filled with randomly oriented actin filaments.
\end{abstract}




\maketitle

\section*{Introduction}
\label{sec:introduction}

\noindent
Intracellular transport of various cargoes from specific origins to target locations is crucial for the correct function of cells and organisms. The cytoskeleton is a self-organizing filamentous network that shapes the mechanical and rheological characteristics of the cell \cite{Nedelec1997,Thoumine1997,Kruse2004,Storm2005,Mizuno2007}, drives cell motility or division \cite{TheCell} and coordinates cargo transport between different cellular regions \cite{Sheetz1983,Howard1989,Appert2015}. The main components of the cytoskeleton which are involved in intracellular transport are the polarized microtubules and actin filaments. 
Associated motor proteins walk actively along these filaments, while they simultaneously bind to cargo particles \cite{Schliwa2003,Vale2000}. 
Kinesin and dynein motors, which run on microtubules, cooperate in transport with myosin motors, which walk on actin filaments. Several motors of diverse species are attached to one cargo concurrently \cite{Balint2013,Hancock2014,Vershinin2007,Welte2004,Gross2007}. Correct cargo delivery thus depends on the coordination of microtubule- and actin-based transport. Experimental data suggests that such a coordination is achieved by regulation of the respective motor species activity through signaling processes \cite{Slepchenko2007,Kural2007,Gross2007,Mallik2004}. A prominent example is the transfer between actin and microtubule network of pigment granules in melanophores which is tightly controlled by the intracellular level of cAMP (cyclic adenosine monophosphate) \cite{Rodionov2003}.\\

\noindent
Furthermore, intracellular transport switches between two modes of motility: ballistic motion along the cytoskeleton is interrupted by effectively stationary states \cite{Bressloff2013}. Cargo-motor-complexes frequently pause due to mechanical constraints by intersection nodes of the cytoskeleton and potentially cycles of detachment and reattachment processes in the crowded cytoplasm, which subsequently leads to a reorientation of the transport direction \cite{Balint2013,Ali2007,Ross2008,Ross2008B,Weiss2004,Feig2016}. 
The run-and-pause behavior of cargo influences the diffusional properties \cite{Hafner2016b} as well as the efficiency of intracellular transport.
\\

\noindent
Typically, cargo, like proteins, vesicles, and other organelles, emerges in one region of the 
cell, but is needed in some other area or has to fuse with a mobile reaction partner. In the absence of a direct connection between searcher and target, the transport is a stochastic process with random alternations between ballistic motion and reorienting arrest states, which is denoted as intermittent search \cite{Benichou2011}.
A particular set of parameters defining this stochastic process, such as the switching rates between ballistic transport and pauses, represents a specific intermittent search strategy.\\

\noindent
The efficiency of a search strategy is commonly evaluated in terms of the mean first passage time that a randomly moving particle needs in order to find a target. 
It has been shown that tuning the search parameters can substantially decrease the mean first passage time in homogeneous and isotropic environments, i.e. under the assumption of a spatially constant density of filament orientations with no preferred direction. For instance, it has been determined that an optimal choice of the switching rates enhances the efficiency of intermittent search strategies \cite{Loverdo2008,Benichou2011,Benichou2005,Loverdo2009} and the search of small targets on the surface of spherical domains also benefits from modulated phases of surface-mediated and bulk diffusion \cite{Benichou2010, Calandre2014}.
Moreover, it has recently been achieved that intracellular transport can profit from inhomogeneous cytoskeletal organizations by demonstrating that the transit time from the nucleus to the whole cell boundary can be reduced by confining the cytoskeleton to a delimited shell within the cell \cite{Ando2015}.\\

\begin{figure*}[t]
\centering
\includegraphics{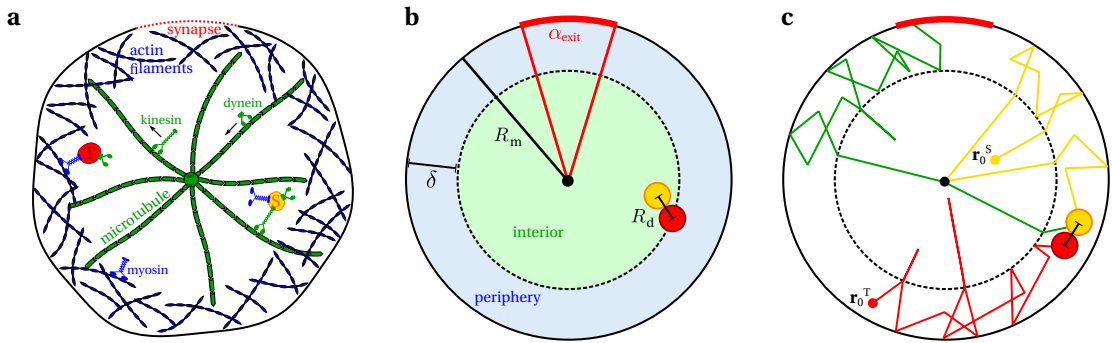}
\caption
{
Random velocity model for intracellular search processes.
\textbf{a} 
Targeted intracellular transport by kinesin, dynein and myosin motor proteins is a stochastic process which strongly depends on the spatial organization of the cytoskeleton.
The cytoskeleton of cells with a centrosome is generally inhomogeneous. While microtubules emanate radially from the central MTOC, actin filaments accumulate in the cortical layer underneath the plasma membrane and exhibit random orientations.
\textbf{b} We idealize the cytoskeleton structure in a spherical cell of radius $R_\text{m}$ by introducing a well-defined actin cortex of width $\delta$ which splits the cytoplasm into interior and periphery. An exit zone on the cell boundary is characterized by the angle $\alpha_\text{exit}$ and the detection distance of two particles is determined by $R_\text{d}$.
\textbf{c} In the combined reaction-escape problem, two mobile particles which are initially located at $\boldsymbol{r}_0^{S}$, $\boldsymbol{r}_0^{T}$ perform a stochastic motion until they detect each other. Following the reaction, the product particle is stochastically transported to the exit zone on the plasma membrane, as sketched by the green trajectory, where the reaction-escape problem is terminated.
}
\label{figure0}
\end{figure*}

\noindent
However, real cell cytoskeletons display a complex spatial organization, which is neither homogeneous nor isotropic. For instance in cells with a centrosome the microtubules emanate radially from the central microtubule organizing center (MTOC) and actin filaments form a thin cortex underneath the plasma membrane with a broad distribution of directions again centered around the radial direction \cite{TheCell}, see Fig. \ref{figure0} a for a sketch. Consequently, the cytoskeleton is very inhomogeneous and characterized by a well defined actin cortex \cite{Eghiaian2015,Salbreux2012}. 
In essence, the specific spatial organization of the cytoskeleton represents, in conjunction with motor-assisted transport, an intermittent search strategy, which is probably optimized for specific frequently occurring transport tasks, but less well suited for others.
It
is still obscure how much the efficiency of diverse transport tasks is effected by the interplay between spatially inhomogeneous cytoskeleton organization, motor performance and detection mode, in particular in comparison to homogeneous search strategies.\\

\noindent
Here, we investigate this question by formulating a random velocity model with intermittent arrest states in spatially inhomogeneous environments.
With the aid of computer simulations we analyze the search in terms of the mean first passage time to detection for three basic tasks, which are frequently encountered in intracellular transport:\\
1) Transport of cargo from an arbitrary position within the cell, typically from a location close to the nucleus, to a specific area on the plasma membrane. For instance directed secretion by immune cells requires the formation of an immunological synapse \cite{Grakoui1999,Bromley2001}, and transport of secretion material towards the synapse involves active motion along the cytoskeleton \cite{Angus2013,Ritter2013}. 
Active transport towards a specific area on the plasma membrane is also required to reinforce spatial asymmetries of proteins which regulate polarized cell functions such as Cdc42 in budding yeast \cite{Marco2007}, to induce the outgrowth of dendrites or axons from neurons \cite{Alberts2003,Chada2003} or to recover plasma membrane damages \cite{McNeil2005,Andrews2014}. The stochastic search for a specific small area on the boundary of a search domain is reminiscent of the so-called narrow escape problem \cite{Schuss2007,Schuss2012}. Here we ask, whether the specific organization of the cytoskeleton, as sketched in Fig. \ref{figure0} a, has the potential to solve the narrow escape problem more efficiently than the homogeneous pendant.\\
2) The enhancement of the reaction kinetics between two reaction partners by motor-assisted transport. 
We consider the binding time of two independently moving particles within the cell, such as fusing vesicles, e.g. late endosomes and lysosomes \cite{TheCell,Granger2014}.
In particular, we check for the impact of an inhomogeneous cytoskeleton organization and two different reaction modes. In the context of intermittent search strategies it is typically assumed that detection is only possible in the phase of slow displacement \cite{Benichou2011,Schwarz2016,Schwarz2016b}, i.e. in the waiting state. Although some experimental data suggest a connection between mobility of particles and likelihood of reactions \cite{Huet2006,Cebecauer2010}, it remains elusive whether this assumption is valid for all chemical reactions which take place inside living cells. Therefore we study two possible detection modes. Reaction may either be possible by simple encounter, no matter what state both particles possess, or it may exclusively be possible when both particles are in the waiting state.\\
3) Finally the combination of the reaction and escape problem, where cargo has first to bind to a reaction partner before it can be delivered or dock at a specific area of the cell boundary, such as a synapse. A prominent realization is the docking of lytic granules at the immunological synapse of 
cytotoxic T-lymphocytes \cite{Grakoui1999,Angus2013,Ritter2013} that requires the pairing with CD3-containing endosomes beforehand \cite{Qu2011}. Lytic granules have a low docking probability at the synapse, whereas endosomes loaded with CD3 receptors have a high docking probability. Apparently, it represents an advantageous strategy to tether lytic granules and CD3 endosomes first in order to guarantee the delivery of cytotoxic cargo exclusively to the synapse. We analyze how the spatial organization of the cytoskeleton supports the efficiency of this strategy for different reaction modes.\\
In the following we refer to these three problems of intracellular transport as 1) the narrow escape problem, 2) the reaction problem, and 3) the reaction-escape problem.\\

\noindent
Our goal is to show that in all cases spatially inhomogeneous intermittent search strategies exist, which are more efficient than their homogeneous counterpart and which are actually realized by the spatial organization of the cytoskeleton of cells with a centrosome, see Fig. \ref{figure0} a.\\

\noindent
In the following we present an extension of \cite{Hafner2016}. There we focused on the narrow escape problem for random walks with intermittent arrest states. Here we study additionaly the reaction as well as the reaction-escape problem in inhomogeneous environments. Our results are in agreement with recent findings for inhomogeneous search strategies with intermittent diffusion in the limit of a vanishing diffusion coefficient \cite{Schwarz2016,Schwarz2016b}. In contrast to \cite{Schwarz2016,Schwarz2016b}, we systematically study the impact of pausing states on the efficiency of intracellular transport task. We further explore the impact of the detection mode on the reaction kinetics, focus on mobile reaction partners with identical as well as nonidentical motility properties and take several steps towards a more realistic distribution of cytoskeleton filaments.

\section*{Methods}
\label{sec:model}

\noindent
For this purpose, we present a coarse grained perspective on intracellular transport by exploring the effective cargo movement, while neglecting the stepping of individual motors at the molecular level.
We formulate an intermittent random velocity model in continuous two-dimensional space and time, in which a particle performs a random walk in two alternating motility modes: (i) a ballistic motion state at velocity $\boldsymbol{v}(v,\phi_v)$, which is associated to directed transport by molecular motors between binding and unbinding events, and (ii) a waiting state in which motors are unbound or particles are stuck at filament crossings. The speed $v$ is assumed to be constant throughout the following. Transitions from motion to waiting state (waiting to motion state) are determined by a constant rate $k_{m\rightarrow w}$ ($k_{w\rightarrow m}$). Hence, the residence times $t_m$, $t_w$ in each state of motility are exponentially distributed
\begin{align}
p(t_m) &= k_{m\rightarrow w} e^{-k_{m\rightarrow w}t_m}, \\
p(t_w) &= k_{w\rightarrow m} e^{-k_{w\rightarrow m}t_w},
\end{align}
with mean values $1/k_{m\rightarrow w}$, $1/k_{w\rightarrow m}$. This is biologically reasonable as active lifetimes of cargoes are reported to be exponentially distributed \cite{Arcizet2008}. While the rate $k_{m\rightarrow w}$ represents the detachment rate of the particle, $k_{w\rightarrow m}$ defines the mean waiting time per arrest state.\\

\noindent
Subsequent to a waiting period, the particle changes its direction of motion according to a characteristic rotation angle $\alpha_\text{rot}$ which reflects the specific spatial organization of the cytoskeleton.
We idealize the cytoskeleton of a spherical cell with radius $R_\text{m}$ by introducing a well-defined actin cortex of width $\delta$ underneath the plasma membrane, which splits the cytoplasm into interior and periphery, as sketched in Fig. \ref{figure0} b. While the interior only contains microtubules, which emanate radially from the central MTOC, the periphery is dominated by randomly oriented actin filaments.
With regard to the spatial organization of the cytoskeleton, the walking direction $\phi_v=\phi_r+\alpha_\text{rot}$ is updated with respect to the polar angle of the current position $\boldsymbol{r}(r,\phi_r)$ and the rotation angle $\alpha_\text{rot}$ is drawn from the idealized distribution
\begin{equation}
f(\alpha_\text{rot})=
\begin{cases}
p_\text{antero}\,\delta(\alpha_\text{rot})\!\!\!\!\!\!&+\,(1{-}p_\text{antero})\,\delta(\alpha_\text{rot}{-}\pi), \\
&\text{ for } 0<r<R_\text{m}{-}\delta,\\
1/(2\pi), & \text{ for } R_\text{m}{-}\delta<r<R_\text{m},
\end{cases}
\label{rotangledistr}
\end{equation}
where $p_\text{antero}$ is the probability for anterograde transport (radially outward transport along microtubule plus direction) and $(1{-}p_\text{antero})$ is the probability for retrograde transport (radially inward transport along microtubule minus direction).
Hence $p_\text{antero}$ reflects the activity level of a particular motor species. For instance $p_\text{antero}{=}0$ corresponds to a high activity of dyneins which carry the cargo to the cell center while for $p_\text{antero}{=}1$ the cargo is mainly transported in the periphery by myosins. Intermediate values of $p_\text{antero}$ enable a frequent exchange between microtubule- and actin-based transport. The distribution of velocity directions given by Eq. \ref{rotangledistr} depends on the current particle position $\boldsymbol{r}$ within the cell which accounts for the inhomogeneity of the cytoskeleton.
The distribution $f(\alpha_\text{rot})$ together with the state transition rates $k_{m\rightarrow w}$, $k_{w\rightarrow m}$ defines a search strategy which is generally inhomogeneous and anisotropic. Here, the special case $\delta{=}R_\text{m}$ leads to a spatially homogeneous and isotropic search strategy.
\\

\noindent
At time $t{=}0$, we assume the particle to start its search in the motion state at position ${\boldsymbol{r}_0}$, which may either be uniformly distributed throughout the cell ($f(r_0){=}1/R_\text{m}$, $f(\phi_{r_0}){=}1/(2\pi)$) or deterministically at the center of the cell ($r{=}0$). Apart from stochastic transitions to the waiting state with rate $k_{m\rightarrow w}$ a ballistically moving particle pauses automatically at the MTOC ($r{=}0$), at the inner border of the actin cortex ($r{=}R_\text{m}{-}\delta$), and at the cell membrane ($r{=}R_\text{m}$), where the rotation angle distribution $f(\alpha_\text{rot})$ is restricted to available values. For the narrow escape problem, the search is terminated when the particle hits the plasma membrane at the exit zone of opening angle $\alpha_\text{exit}$ as illustrated in Fig. \ref{figure0} b. In the case of the reaction problem, the search is terminated by encounter of searcher and target particle ($|\boldsymbol{r}^\text{S}{-}\boldsymbol{r}^\text{T}|{=}R_\text{d}$, see Fig. \ref{figure0} b and c) either regardless of their motility state or only possible in the waiting state. Reaction takes place instantaneously.
In the reaction-escape problem searcher and target particle first have to react before the newly formed product particle can be transported to a specific zone on the membrane of the cell.
In the following we use rescaled, dimensionless spatial and temporal coordinates
$\boldsymbol{r}\!\mapsto\!\boldsymbol{r}/R_\text{m}$, $t\!\mapsto\! vt/R_\text{m}$ 
and parameters $k_{m\rightarrow w}\!\!\mapsto\!\!\! R_\text{m}k_{m\rightarrow w}/v$, $k_{w\rightarrow m}\!\mapsto\! R_\text{m}k_{w\rightarrow m}/v$, $\delta\!\mapsto\!\delta/R_\text{m}$.\\

\noindent
The efficiency of a search strategy, defined by a specific cytoskeleton organization and motor performance, is measured in terms of a mean first passage time (MFPT) with respect to the events defined by the different search problems.
To calculate the MFPT we use an event-driven Monte Carlo algorithm to generate the stochastic process, as sketched in Fig. \ref{figure0} c, and calculate the MFPT as an ensemble average. We apply of the order of $10^6$ independent realizations of the search process for each parameter value such that the relative statistical error for the MFPT is significantly lower than $0{.}5\%$.

\begin{figure*}[t]
\centering
\includegraphics{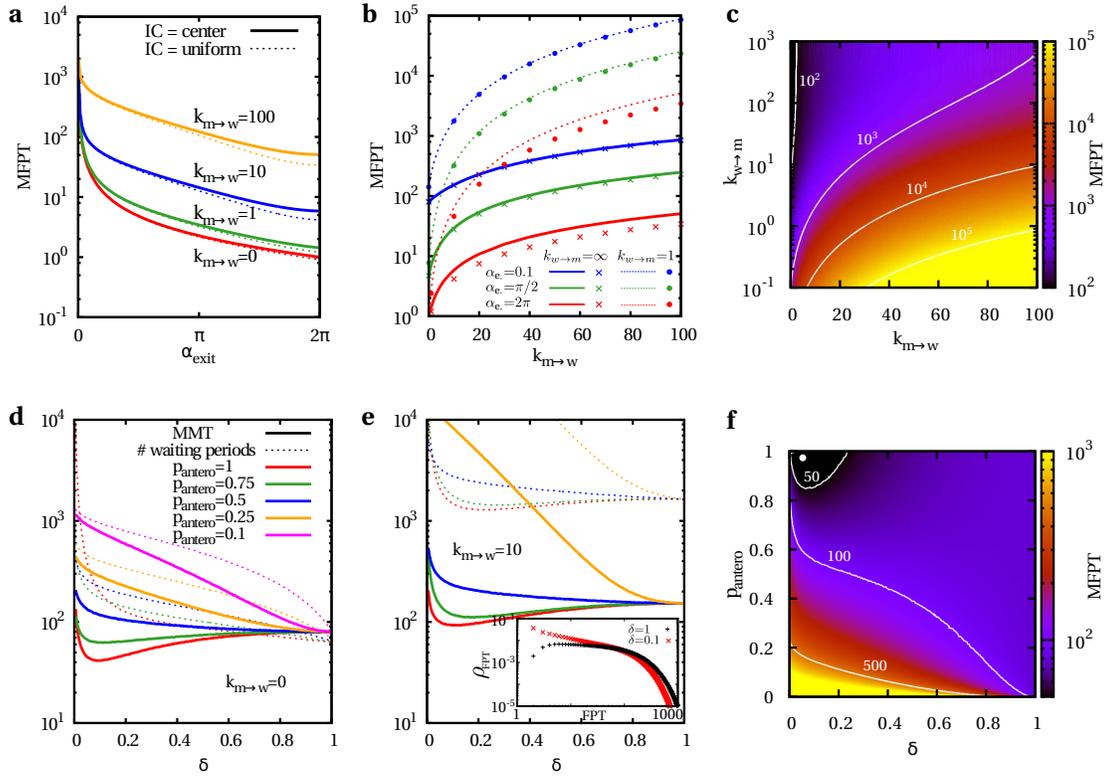}
\caption
{Narrow escape problem for homogeneous and inhomogeneous spatial organizations of the cytoskeleton.
\textbf{a} The MFPT in dependence of the target size $\alpha_\text{exit}$ for different values of the transition rate $k_{m\rightarrow w}$ and $k_{w\rightarrow m}{=}\infty$ on a homogeneous cytoskeleton structure. Dashed lines correspond to initially uniformly distributed searcher positions, whereas solid lines refer to searchers which are initially located at the center of the cell.
\textbf{b} The MFPT versus $k_{m\rightarrow w}$ for a homogeneous cytoskeleton organization and diverse target sizes $\alpha_\text{exit}$ and rates $k_{w\rightarrow m}$. Lines are associated to the centered initial condition, while symbols correspond to the uniform case.
\textbf{c} The MFPT for a homogeneous cytoskeleton organization and centered initial condition in dependence of the transition rates $k_{m\rightarrow w}$ and $k_{w\rightarrow m}$ for $\alpha_\text{exit}{=}0{.}1$.
\textbf{d} The MMT and the $\#$ waiting periods against the width of the actin cortex $\delta$, which defines the inhomogeneity of the cytoskeleton, for diverse $p_\text{antero}$, $k_{m\rightarrow w}{=}0$, centered initial condition and exit size $\alpha_\text{exit}{=}0{.}1$.
\textbf{e} The same as in d, but for $k_{m\rightarrow w}{=}10$. The inset shows the full distribution of first passage times for $k_{m\rightarrow w}{=}10$, $k_{w\rightarrow m}{=}\infty$, $p_\text{antero}{=}1$, centered initial condition and $\delta{=}1$, $\delta{=}0{.}1$. At intermediate timescales a broadening of the power law regime is observed for $\delta\to 0$.
\textbf{f} The MFPT versus $\delta$ and $p_\text{antero}$ for $k_{m\rightarrow w}^\text{opt}{=}0$, $k_{w\rightarrow m}^\text{opt}{=}\infty$, which constitutes the optimal choice of transition rates for a spatially homogeneous cytoskeleton, and centered initial condition of the searcher in the case of $\alpha_\text{exit}{=}0{.}1$.
}
\label{figure1}
\end{figure*}

\section*{Results}

\noindent
The proposed model allows the study of diverse transport tasks. Here we focus on three different, biologically relevant search problems: the narrow escape, the reaction and the reaction-escape problem.
We analyze the dependence of the search efficiency on the spatial organization of the cytoskeleton as well as the detection mode and the motor performance. 

\begin{figure*}[t]
\centering
\includegraphics{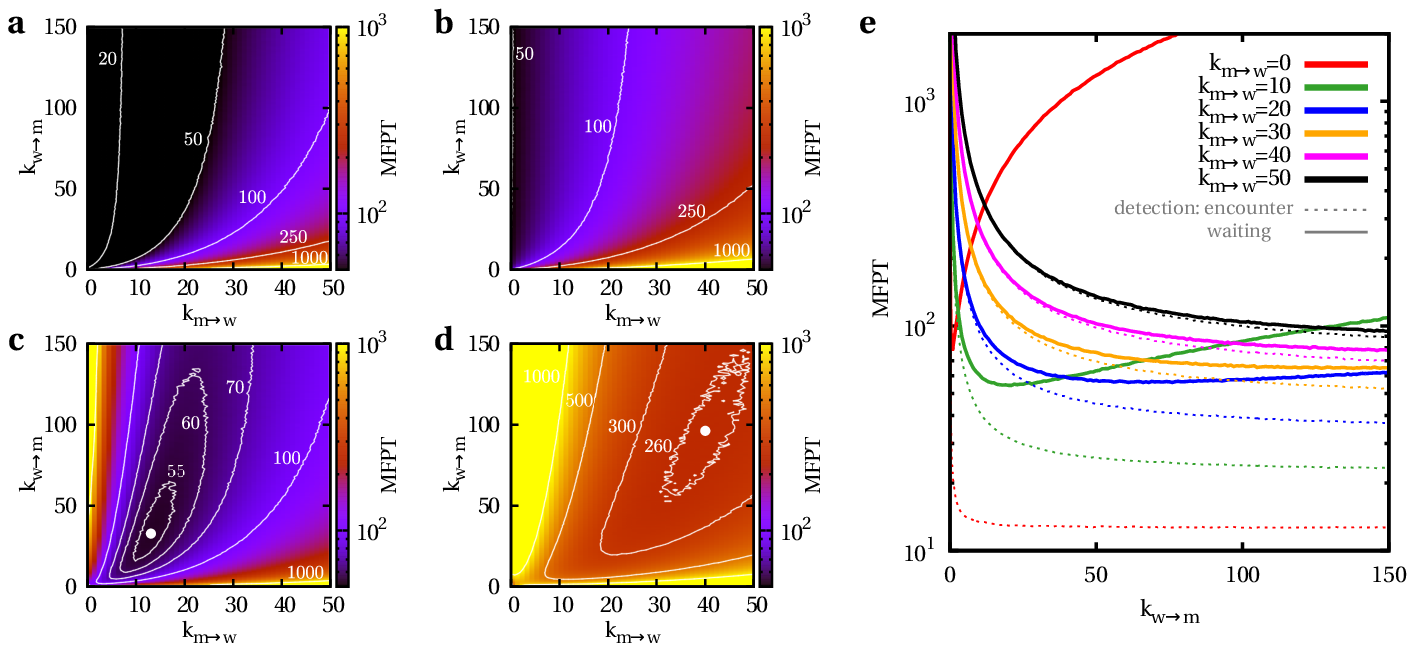}
\caption
{Reaction problem for homogeneous organizations of the cytoskeleton.
\textbf{a} The MFPT versus $k_{m\rightarrow w}$ and $k_{w\rightarrow m}$ for $R_\text{d}{=}0{.}1$ and detection by simple encounter.
\textbf{b} The same as in a, but for $R_\text{d}{=}0{.}025$.
\textbf{c} The MFPT versus $k_{m\rightarrow w}$ and $k_{w\rightarrow m}$ for $R_\text{d}{=}0{.}1$ and detection which is only possible in the waiting state.
\textbf{d} The same as in c, but for $R_\text{d}{=}0{.}025$.
\textbf{e} The MFPT in dependence of $k_{w\rightarrow m}$ for diverse rates $k_{m\rightarrow w}$, $R_\text{d}{=}0{.}1$ and both detection modes.
}
\label{figure2}
\end{figure*}

\subsection*{Narrow Escape Problem}

\noindent
First we consider intracellular transport of cargo which is initially either centered or uniformly distributed within the cell to a specific area on the the plasma membrane. 
We evaluate the impact of a particular spatial organization of the cytoskeleton, given by Eq. \ref{rotangledistr}, on the search efficiency.
\\

\noindent
In order to demonstrate the gain in efficiency of a spatially inhomogeneous search strategy, corresponding to $0<\delta<1$, we first consider homogeneous cytoskeleton organizations with $\delta{=}1$. 
In accordance to \cite{Hafner2016}, we find that the MFPT decreases monotonically with increasing target size $\alpha_\text{exit}$ for diverse values of the transition rates $k_{m\rightarrow w}$ and $k_{w\rightarrow m}$, as shown in Fig. \ref{figure1} a and b. 
Fig. \ref{figure1} a and b further display, that initially uniformly distributed searcher positions are slightly more advantageous than a deterministic initial position at the cell center for large target sizes. For great exit zones $\alpha_\text{exit}\approx 2\pi$, the searcher is likely to find the escape by first encounter with the membrane, hence the initial location of the searcher has a strong influence on the MFPT. Contrarily, for narrow escapes many hitting events with the membrane are necessary to find the exit zone and, as a result, the impact of the initial position of the searcher vanishes, as the findings in Fig. \ref{figure1} a and b demonstrate.
Moreover, Fig. \ref{figure1} a and b indicate that an increase in the rate $k_{m\rightarrow w}$ as well as a decrease in $k_{w\rightarrow m}$ hinders the detection of targets alongside the membrane. Accordingly, Fig. \ref{figure1} c illustrates that the optimal choice of the transition rates for a search on a homogeneous cytoskeleton is given by $k_{m\rightarrow w}^\text{opt}{=}0$ and $k_{w\rightarrow m}^\text{opt}{=}\infty$. Consequently, an uninterrupted motion pattern without directional changes in the bulk of the cell constitutes an optimal search strategy for a homogeneous cytoskeleton ($\delta{=}1$). In the following, we ask whether an inhomogeneous filament structure ($\delta\!<\!1$) has the potential to solve the narrow escape problem more efficiently than its homogeneous pendant.\\

\noindent
To answer this question, we compute the influence of the actin cortex width $\delta$ on the MFPT for various parameters $k_{m\rightarrow w}$, $k_{w\rightarrow m}$, and $p_\text{antero}$ for small escape regions of angle $\alpha_\text{exit}{=}0{.}1$. Note that in this case the exit zone constitutes only $1.59\%$ of the total spherical surface. 
Remarkably, we find that for large probabilities of anterograde transport $p_\text{antero}$ the MFPT exhibits a pronounced minimum at small widths of the actin cortex $\delta$. This phenomenon is largely robust against changes in the transition rates $k_{m\rightarrow w}$ and $k_{w\rightarrow m}$. The MFPT can be split into the total mean motion time (MMT) and the total mean waiting time (MWT) a particle experiences in the course of its search. Thereby, the MWT is the product of the total mean number of waiting periods ($\#$ waiting periods) and the mean waiting time per arrest state ($1/k_{w\rightarrow m}$). In \cite{Hafner2016}, we showed that the number of waiting periods is independent of $k_{w\rightarrow m}$ and defines the optimal width of the actin cortex $\delta$ in the limit of $k_{w\rightarrow m}\to 0$. Consistent with \cite{Hafner2016}, Fig. \ref{figure1} d and e demonstrate that the MMT, which defines the MFPT for $k_{w\rightarrow m}\to\infty$, as well as the mean number of waiting periods display prominent minima in dependence of $\delta$ for large probabilities $p_\text{antero}$ for anterograde transport.
Note that the first passage time distribution which for Brownian motion in bounded domains generally consists of three parts (short times: exponential decay, intermediate times: power law, long times: exponential decay) \cite{Schehr2011,Oshanin2012} exhibits a broadening of the intermediate power law regime for $\delta \to 0$, as shown in the inset of Fig. \ref{figure1} e. However, when the search is dominated by radially inward transport for small $p_\text{antero}$, a homogeneous strategy $\delta{=}1$ is most efficient, since $\delta\!\neq\! 1$ and $p_\text{antero}{=}0$ prevents the searcher from touching the membrane. Nonetheless, Fig. \ref{figure1} f shows that a small cortex width $\delta$ potentially reduces the MFPT by several order of magnitude. This emphasizes the general enhancement of the search efficiency by a spatially inhomogeneous filament structure compared to the homogeneous pendant. 

\subsection*{Reaction Problem}

\begin{figure*}[t]
\centering
\includegraphics{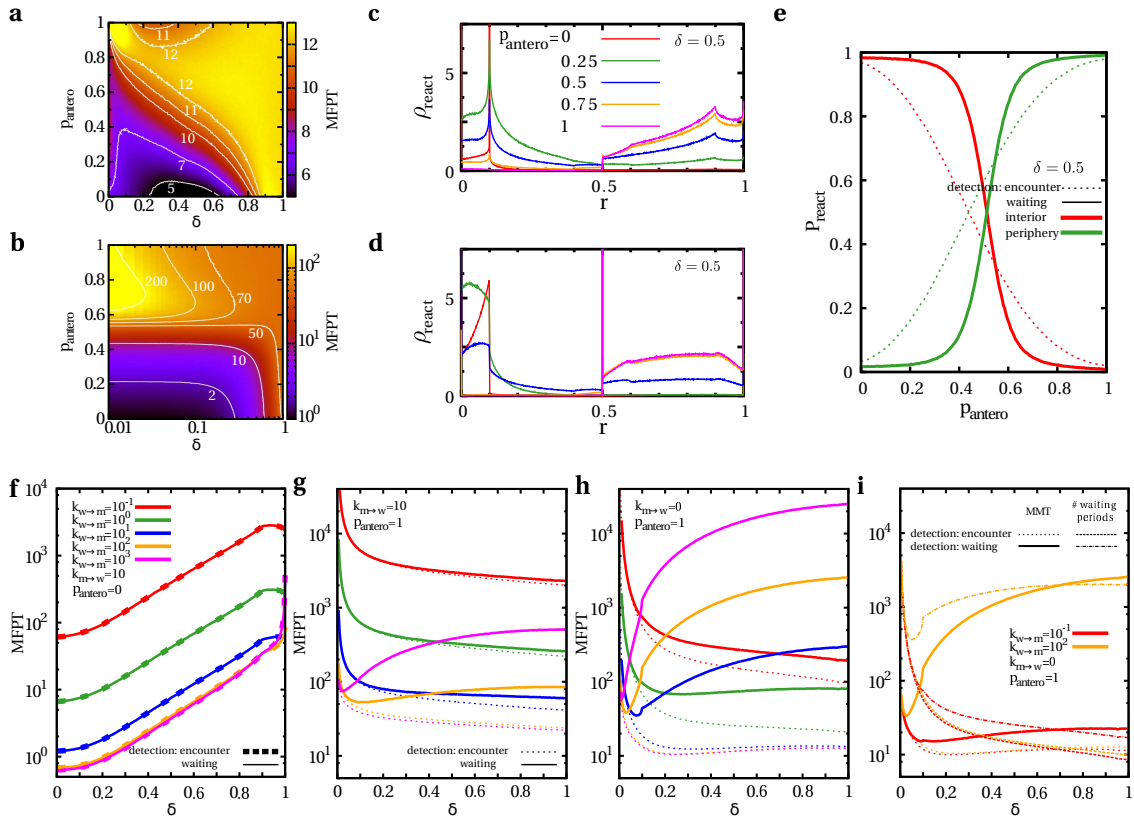}
\caption
{
Reaction problem for inhomogeneous organizations of the cytoskeleton.
\textbf{a} 
The MFPT in dependence of $\delta$ and $p_\text{antero}$ for $R_\text{d}{=}0{.}1$, $k_{m\rightarrow w}^\text{opt}{=}0$ and $k_{w\rightarrow m}^\text{opt}{=}\infty$ which are the optimal rates for the homogeneous case when detection happens by simple encounter.
\textbf{b} 
The MFPT in dependence of $\delta$ and $p_\text{antero}$ for $R_\text{d}{=}0{.}1$, $k_{m\rightarrow w}^\text{opt}{=}13$ and $k_{m\rightarrow w}^\text{opt}{=}34$ which are the optimal rates for the homogeneous case, since detection is only possible in the waiting state.
\textbf{c} 
Distribution of reaction locations $r$ for various values of $p_\text{antero}$. The cortical width is fixed to $\delta{=}0{.}5$ and the transition rates $k_{m\rightarrow w}^\text{opt}$, $k_{w\rightarrow m}^\text{opt}$ are applied. Detection takes place by encounter at $R_\text{d}{=}0{.}1$.
\textbf{d} 
The same as in c but detection is only possible in the waiting state.
\textbf{e} 
Probability for a reaction to take place in the interior or the periphery of a cell versus $p_\text{antero}$ for fixed width of the actin cortex $\delta{=}0{.}5$ and $R_\text{d}{=}0{.}1$. The rates $k_{m\rightarrow w}^\text{opt}$, $k_{w\rightarrow m}^\text{opt}$ are applied for both reaction modes.
\textbf{f} 
The MFPT in dependence of width of the actin cortex $\delta$ for $R_\text{d}{=}0{.}1$, $p_\text{antero}{=}0$, $k_{m\rightarrow w}{=}10$, diverse values of $k_{w\rightarrow m}$ and both detection modes.
\textbf{g} 
The same as in f but for $p_\text{antero}{=}1$.
\textbf{h}
The same as in g but for $k_{m\rightarrow w}{=}0$.
\textbf{i} 
The MMT as well as the mean number of waiting periods versus $\delta$ for $R_\text{d}{=}0{.}1$, $p_\text{antero}{=}1$, $k_{m\rightarrow w}{=}0$, diverse values of $k_{w\rightarrow m}$ and both detection modes.
}
\label{figure3}
\end{figure*}

\noindent
Next, we investigate how efficient a spatial organization of the cytoskeleton as defined in Eq. \ref{rotangledistr} can actually be for the reaction problem.
Two mobile particles with identical properties and uniformly distributed initial positions perform random motion inside the cell until they detect each other and react. We consider two different reaction modes which depend on the motility states of both particles. On the one hand, searcher and target particle react by simple encounter $|\boldsymbol{r}^\text{S}-\boldsymbol{r}^\text{T}|{=} R_\text{d}$ regardless of the state both are in. On the other hand, detection is only possible when searcher and target particle are in the waiting state and their relative distance is smaller than the detection radius $|\boldsymbol{r}^\text{S}-\boldsymbol{r}^\text{T}|\leq R_\text{d}$. Fig. \ref{figure0} c includes a sketch of the reaction problem.\\

\noindent
Fig. \ref{figure2} a-d show the results for the MFPT in the case of a spatially homogeneous 
cytoskeleton ($\delta{=}1$) as a function of the transition rates $k_{m\rightarrow w}$ and $k_{w\rightarrow m}$ for two different detection radii $R_\text{d}{=}0{.}1$ and $R_\text{d}{=}0{.}025$. While Fig. \ref{figure2} a,b display that the optimal choice of transition rates is $k_{m\rightarrow w}^\text{opt}{=}0$, $k_{w\rightarrow m}^\text{opt}{=}\infty$ for detection by simple encounter, a non-trivial optimum arises when reaction is only possible in the waiting state, as shown in Fig. \ref{figure2} c,d. Remarkably, the absolute values of the minima are rather robust against alterations of the transition rates, as indicated by the contour lines. Moreover, Fig. \ref{figure2} e illustrates that for high values of $k_{m\rightarrow w}$ the MFPT is largely independent of the detection mode. The detection time is then dominated by getting the particles in close proximity in the first place.\\
 
\noindent
We take the optimal values $k_{m\rightarrow w}^\text{opt}$, $k_{w\rightarrow m}^\text{opt}$
from the homogeneous case $\delta{=}1$ and compute the MFPT for an inhomogeneous cytoskeleton organization in dependence of the actin cortex width $\delta$ and the probability for radially outward transport $p_\text{antero}$. The results, shown in Fig. \ref{figure3} a and b, demonstrate that for both detection modes a superior strategy can be found for $p_\text{antero}{=}0$, which drives the particles towards the central MTOC.
Fig. \ref{figure3} c-e display that a cell is able to regulate the location of the reactions, and thus the spatial distribution of reaction products, by adjustment of $p_\text{antero}$. While for $p_\text{antero}{=}0$ reaction mainly takes place in the interior of the cell close to the central MTOC, it predominantly happens in the peripheral actin cortex for $p_\text{antero}{=}1$.
Note that the peaks in Fig. \ref{figure3} c and d at radial positions $r{=}R_\text{d}$ and $r{=}R_\text{m}-R_\text{d}$ arise due to the detection radius which is here set to $R_\text{d}{=}0{.}1$. Furthermore, the peaks at $r\in\{0;\delta;R_\text{m}\}$ result from automatic switching to the arrest state which is particularly beneficial when detection is only possible in the waiting state.
Besides the reaction strategy of moving searcher and target particle towards the MTOC by $p_\text{antero}{=}0$, which is generally optimal for $\delta{=}0$ as indicated by Fig. \ref{figure3} f, the reaction time can also be minimized for $p_\text{antero}{=}1$ by establishing a thin actin cortex, as shown in Fig. \ref{figure3} g and h.
The occurrence of the minimum depends critically on both transition rates. Fig. \ref{figure3} i displays that the MMT as well as the number of waiting periods depend on $k_{m\rightarrow w}$ and $k_{w\rightarrow m}$. This is in sharp contrast to our findings for the narrow escape problem and holds in particular when detection is only possible in the waiting state.
\\

\noindent
Note that for instance the MFPT in Fig. \ref{figure3} h and i displays a prominent kink at $\delta{=}0{.}1$ for $k_{m\rightarrow w}{=}0$ for detection in the waiting state.
Under these conditions, a particle does not switch to the waiting state in the bulk of the cell and consequently detection critically depends upon automatic transitions to the waiting state at the MTOC ($r{=}0$), the inner border of the actin cortex ($r{=}1{-}\delta$), and the plasma membrane ($r{=}1$). When $\delta\leq R_\text{d}$, there are three different realizations of searcher and target locations which may lead to detection: either both particles at the membrane, both particles at the inner border or one particle at the membrane and one at the inner border. When $1{-}R_\text{d}\leq\delta$, there are again three possible configurations for detection: both particles at the inner border, both at the MTOC, one at the inner border and one at the MTOC. However, for $R_\text{d} < \delta < 1{-}R_\text{d}$, the mixed configurations are not possible anymore and reaction can only take place, when both particles are either closely enough located at the membrane or at the MTOC. This results in the kink of the MFPT at $\delta{=}0{.}1$, since $R_\text{d}{=}0{.}1$ in Fig. \ref{figure3}.

\begin{figure*}[t]
\centering
\includegraphics{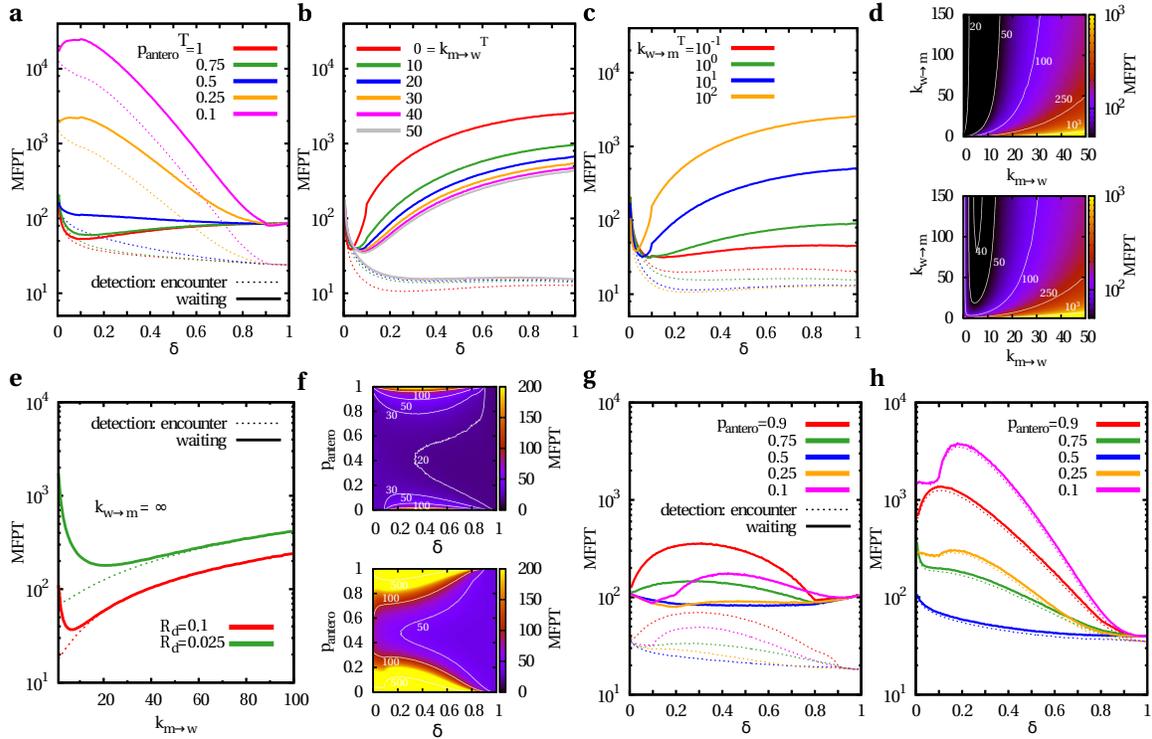}
\caption
{
Reaction problem for nonidentical particles.
\textbf{a}
The MFPT versus $\delta$ for nonidentical values of $p_\text{antero}$. While $p_\text{antero}^\text{S}{=}1$ is fixed for the searcher, $p_\text{antero}^\text{T}$ is varied for the target particle. Apart from that the particles are identical, i.e. $k_{m\rightarrow w}{=}10$, $k_{w\rightarrow m}{=}100$. Both detection modes are considered for $R_\text{d}{=}0{.}1$.
\textbf{b}
The MFPT in dependence of $\delta$ for nonidentical transition rates $k_{m\rightarrow w}$. The rate $k_{m\rightarrow w}^\text{T}$ is varied for the target, whereas the one of the searcher is fixed to $k_{m\rightarrow w}^\text{S}{=}0$. Other parameters are identical, i.e. $k_{w\rightarrow m}{=}100$, $p_\text{antero}{=}1$ and $R_\text{d}{=}0{.}1$.
\textbf{c}
The MFPT versus $\delta$ for nonidentical transition rates $k_{w\rightarrow m}$. The rate $k_{w\rightarrow m}^\text{T}$ of the target particle is varied, while the one of the searcher is given by $k_{w\rightarrow m}^\text{S}{=}100$, otherwise $k_{m\rightarrow w}{=}0$, $p_\text{antero}{=}1$ and $R_\text{d}{=}0{.}1$.
\textbf{d}
The reaction problem for an immobile, uniformly distributed target particle and a homogeneous cytoskeleton in dependence of the transition rates $k_{m\rightarrow w}$ and $k_{w\rightarrow m}$ of the searcher for $R_\text{d}{=}0{.}1$. The upper panel shows the results when detection takes place by encounter, while for the lower one detection depends on the waiting state.
\textbf{e}
The MFPT versus the transition rate $k_{m\rightarrow w}$ for an immobilized target and homogeneous cytoskeleton, where $k_{w\rightarrow m}{=}\infty$ and $R_\text{d}\in\{0{.}025;0{.}1\}$. Both detection modes are applied. 
\textbf{f}
The MFPT versus $\delta$ and $p_\text{antero}$ for an immobile target and $R_\text{d}{=}0{.}1$. The optimal transition rates for the homogeneous pendant are applied for both detection modes, i.e. (upper panel) $k_{m\rightarrow w}^\text{opt}{=}0$, $k_{w\rightarrow m}^\text{opt}{=}\infty$ for detection by encounter and (lower panel) $k_{m\rightarrow w}^\text{opt}{=}7$, $k_{w\rightarrow m}^\text{opt}{=}\infty$ when detection is only possible in the waiting state.
\textbf{g}
The MFPT in dependence of $\delta$ for an immobile target and various $p_\text{antero}$. Both detection modes are considered for $R_\text{d}{=}0{.}1$, $k_{m\rightarrow w}{=}1$ and $k_{w\rightarrow m}{=}\infty$.
\textbf{h}
The same as in g, but for $k_{m\rightarrow w}{=}10$.
}
\label{figure4}
\end{figure*}

\subsection*{Reaction Problem for nonidentical Particles}

\noindent
Until now, we studied the reaction problem for searcher and target particles with identical motility pattern, i.e. with equal transition rates $k_{m\rightarrow w}$, $k_{w\rightarrow m}$ and equal probability for anterograde transport $p_\text{antero}$. Here, we investigate consequences of nonidentical motility schemes on the efficiency of intracellular reactions. 
This is relevant when searcher and target particle are not equal but e.g. are equipped with different motor species or differ in size. Waiting times as well as the effective mesh size of the cytoskeleton (which are associated with the transition rates $k_{m\rightarrow w}$, $k_{w\rightarrow m}$) depend on the size of the transported cargo \cite{Balint2013}.\\

\noindent
In order to study the impact of nonidentical particle properties, we fix the motility parameters ($k_{m\rightarrow w}^\text{S}$, $k_{w\rightarrow m}^\text{S}$, $p_\text{antero}^\text{S}$) of the searching particle, while varying the ones ($k_{m\rightarrow w}^\text{T}$, $k_{w\rightarrow m}^\text{T}$, $p_\text{antero}^\text{T}$) of the target.
Fig. \ref{figure4} a displays the MFPT in dependence of the width $\delta$ of the actin cortex for nonidentical values of $p_\text{antero}$. While $p_\text{antero}^\text{S}{=}1$ is fixed for the searcher, $p_\text{antero}^\text{T}$ is varied for the target particle. Apart from that the particles are identical. As expected, a strong inequality in $p_\text{antero}$ increases the MFPT by several orders of magnitude as it drives the particles apart. While the ones with $p_\text{antero}{=}1$ are predominantly moving in the cortex, the ones with $p_\text{antero}{=}0$ stick to the center of the cell. Consequently, a homogeneous search strategy on a random network with $\delta{=}1$ is favorable in that case. This finding is robust against changes in the transition rates $k_{m\rightarrow w}$, $k_{w\rightarrow m}$.
Remarkably, the influence of nonidentical transition rates $k_{m\rightarrow w}$, $k_{w\rightarrow m}$ on the search efficiency strongly depends on the detection mode, as found by Fig. \ref{figure4} b and c. While making the target less motile than the searcher, either by increasing $k_{m\rightarrow w}^\text{T}$ or decreasing $k_{w\rightarrow m}^\text{T}$, makes the search less efficient when detection takes place by pure encounter, it has the opposite effect when detection is only possible in the waiting state. This result is robust against changes in $p_\text{antero}$ but vanishes for increasing $k_{m\rightarrow w}$, as the search is then first of all determined by getting the particles in close proximity.\\

\noindent
Next, we analyze the consequences of an immotile target which is uniformly distributed in the cell on the efficiency of the reaction problem.
Fig. \ref{figure4} d shows the MFPT to an inert target in the case of a homogeneous cytoskeleton in dependence of the transition rates $k_{m\rightarrow w}$ and $k_{w\rightarrow m}$ of the searcher for $R_\text{d}{=}0{.}1$. The upper panel displays the MFPT for detection by encounter, whereas detection depends on the waiting state in the lower panel. We find that $k_{w\rightarrow m}^\text{opt}{=}\infty$ is optimal for both detection modes and all values of the detection radius $R_\text{d}$. Moreover, $k_{m\rightarrow w}^\text{opt}{=}0$ is universally optimal for detection by encounter, whereas $k_{m\rightarrow w}^\text{opt}$ depends on $R_\text{d}$ when detection is only possible in the waiting state, as demonstrated by Fig. \ref{figure4} e.
In order to investigate the impact of inhomogeneous cytoskeleton organizations, we evaluate the MFPT to an immobile target in Fig. \ref{figure4} f in dependence of $p_\text{antero}$ and $\delta$ for both detection modes, where the optimal transition rates for the homogeneous counterparts are applied. For all widths $\delta$ of the actin cortex, an optimal strategy to detect an immobile target within the cell is defined by $p_\text{antero}{=}0{.}5$. In general, a homogeneous, random cytoskeletal network ($\delta{=}1$, independent of $p_\text{antero}$) is most efficient. This conclusion is very robust against changes in the transition rates $k_{m\rightarrow w}$, $k_{w\rightarrow m}$ and holds for both detection modes, as indicated by Fig. \ref{figure4} g and h.
While the reaction problem for two identically motile particles can be efficiently solved by restricting motility space (either by $p_\text{antero}{=}0$ and $\delta{=}0$ or by $p_\text{antero}{=}1$ and a thin cortex $\delta$, see Fig. \ref{figure3}), it is best to explore the cell on a homogeneous network ($\delta{=}1$ or at least $p_\text{antero}{=}0{.5}$) in the case of randomly distributed immotile targets.

\subsection*{Reaction-Escape Problem}

\noindent
Next we study the efficiency of inhomogeneous search strategies for the combination of reaction and escape problem. Cargo first has to bind to a reaction partner before the product can be delivered to a specific area on the cell boundary. All particles obey the same motility scheme. 
Fig. \ref{figure0} c shows a sketch of the reaction-escape problem.
The searcher and the target particle react and form a product particle once they get closer than a distance $R_\text{d}$ either regardless of their motility state or only possible in the waiting state. Then, the product particle performs the escape problem, as the exit area is only absorbing for the product. The total MFPT of the reaction-escape problem is composed of the MFPT$_\text{react}$ for the reaction problem and the MFPT$_\text{escape}$ for the following escape problem of the product particle to the exit zone on the plasma membrane, i.e. $\text{MFPT}{=} \text{MFPT}_\text{react} {+} \text{MFPT}_\text{escape}$.\\

\begin{figure*}[t]
\centering
\includegraphics{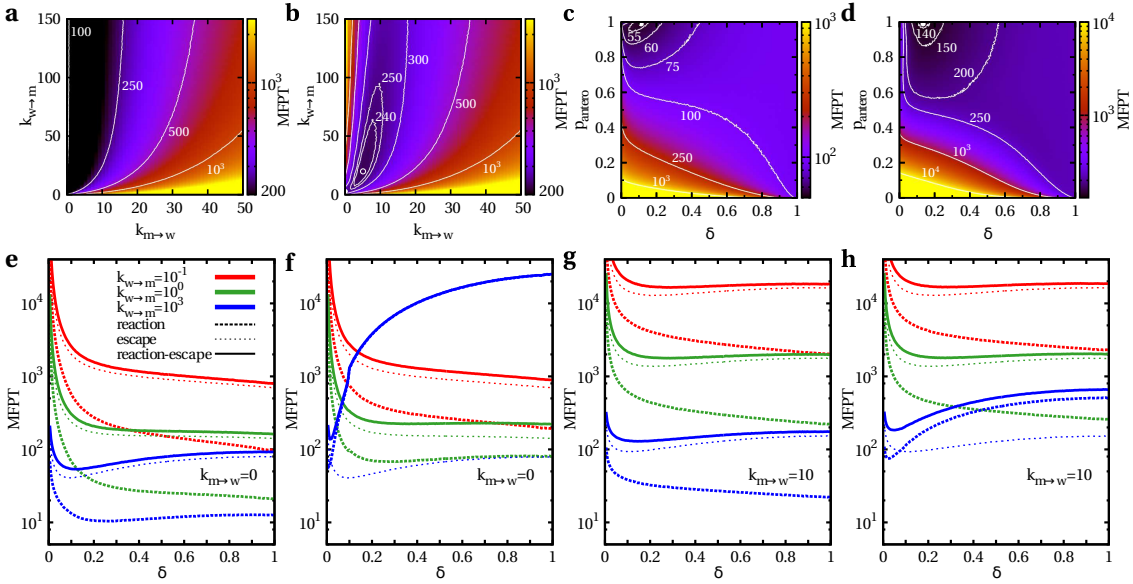}
\caption
{Reaction-escape problem for homogeneous and inhomogeneous spatial organizations of the cytoskeleton.
\textbf{a} The MFPT versus the transition rates $k_{m\rightarrow w}$ and $k_{w\rightarrow m}$ for a homogeneous cytoskeleton organization with reaction radius $R_\text{d}{=}0{.}1$ and exit size $\alpha_\text{exit}{=}0{.}1$, where reaction happens by encounter regardless of the motility state.
\textbf{b} The same as in a, but reaction is only possible in the waiting state.
\textbf{c} The MFPT in dependence of the actin cortex width $\delta$ and the probability for anterograde transport $p_\text{antero}$ for $k_{m\rightarrow w}^\text{opt}{=}0$, $k_{m\rightarrow w}^\text{opt}{=}\infty$, $R_\text{d}{=}0{.}1$ and $\alpha_\text{exit}{=}0{.}1$. Reaction happens by pure encounter.
\textbf{d} The same as in c, but for $k_{m\rightarrow w}^\text{opt}{=}5$ and $k_{w\rightarrow m}^\text{opt}{=}22$. Reaction is only possible when both particles are in the waiting state.
\textbf{e} The MFPT against the width of the actin cortex $\delta$ for $k_{m\rightarrow w}{=}0$ and diverse values of $k_{w\rightarrow m}$, where $p_\text{antero}{=}1$, $R_\text{d}{=}0{.}1$ and $\alpha_\text{exit}{=}0{.}1$. Reaction happens by simple encounter.
\textbf{f} The same as in e but reaction depends on the waiting state.
\textbf{g} The same as in a but for $k_{m\rightarrow w}{=}10$.
\textbf{h} The same as in f but for $k_{m\rightarrow w}{=}10$.
}
\label{figure5}
\end{figure*}

\noindent
Fig. \ref{figure5} a and b present the MFPT for the reaction-escape problem in the case of a homogeneous cytoskeleton ($\delta{=}1$) as a function of the rates $k_{m\rightarrow w}$, $k_{w\rightarrow m}$ for $R_\text{d}{=}0{.}1$ and $\alpha_\text{exit}{=}0{.}1$.
As expected, when reaction does not depend on the motility state of the particles, the MFPT is minimal for $k_{m\rightarrow w}^\text{opt}{=}0$ and $k_{w\rightarrow m}^\text{opt}{=}\infty$, since the MFPT of the homogeneous escape problem and the one of the homogeneous reaction problem is optimal for the same rates, as displayed in Fig. \ref{figure1} c and \ref{figure2} a. 
However, when reaction is only possible in the waiting state, the MFPT features two minima, as a result of the interplay between $k_{m\rightarrow w}{=}0$, $k_{w\rightarrow m}{=}\infty$ being optimal for the escape and $k_{m\rightarrow w}{=}13$, $k_{w\rightarrow m}{=}34$ being optimal for the reaction problem. While the global minimum at $k_{m\rightarrow w}{=}0$, $k_{w\rightarrow m}{=}1$ is deficient, the local minimum at $k_{m\rightarrow w}^\text{opt}{=}5$, $k_{w\rightarrow m}^\text{opt}{=}22$, whose absolute value is only $5\%$ higher than the one of the global minimum, is very pronounced and robust against alterations of the rates.\\

\noindent
In order to explore the influence of the spatial inhomogeneity of the cytoskeleton, we measure the MFPT for the reaction-escape problem as a function of the cortex width $\delta$ and the probability for anterograde transport $p_\text{antero}$ in Fig. \ref{figure5} c and d for both reaction modes, where we used the optimal transition rates $k_{m\rightarrow w}^\text{opt}$ and $k_{w\rightarrow m}^\text{opt}$ of the homogeneous case. 
Even though $p_\text{antero}{=}0$ is advantageous for the pure reaction problem, as found in Fig. \ref{figure3}, a low probability of radially outward transport is highly inferior for the reaction-escape problem as it prevents the compound particle from reaching the membrane and thus from detecting the exit zone. A superior search strategy for the reaction-escape problem is defined by a high probability $p_\text{antero}$ and a thin actin cortex $\delta$. The gain in search efficiency by an inhomogeneous spatial organization of the cytoskeleton is conserved also for non-optimal transition rates $k_{m\rightarrow w}$ and $k_{w\rightarrow m}$, as illustrated in Fig. \ref{figure5} e-h. Consequently, adjustments of the width of the actin cortex potentially reduces the MFPT up to several orders of magnitude.

\begin{figure*}[t]
\centering
\includegraphics{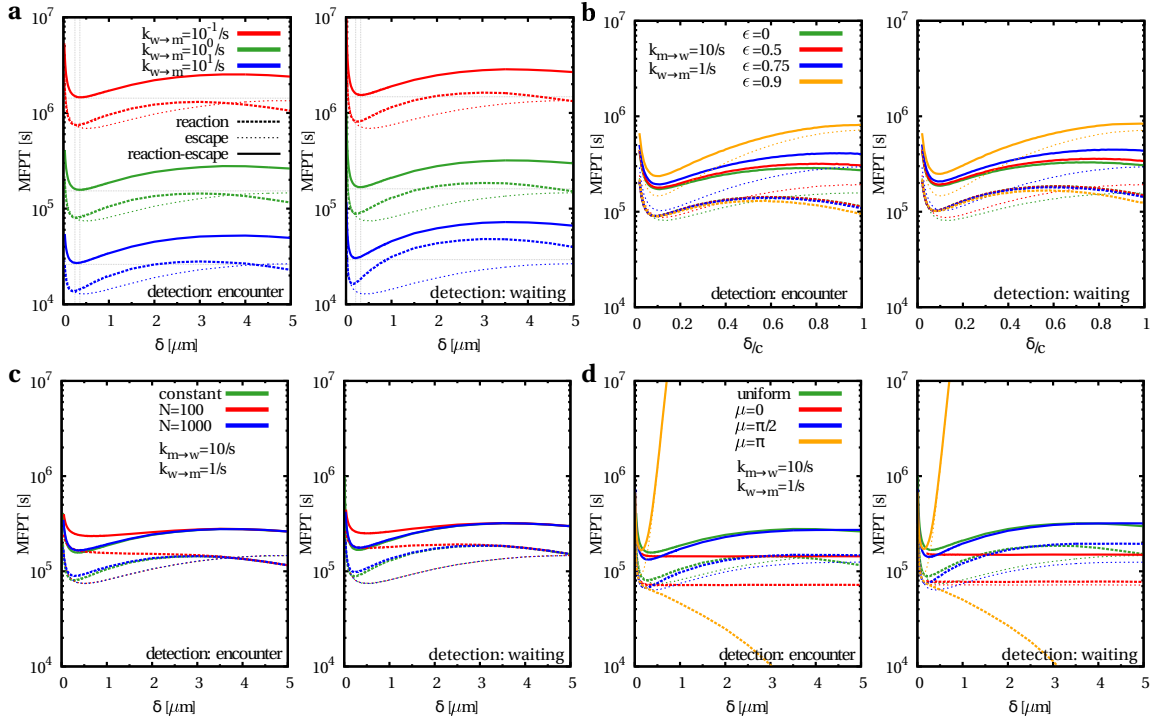}
\caption
{
Steps towards a more realistic cytoskeleton. The MFPT of the escape, the reaction and the reaction-escape problem in dependence of the actin cortex width $\delta$ for a three-dimensional cell of volume $V{=}4\pi(5\,\mu\text{m})^3/3$ with $\alpha_\text{exit}{=}0{.}2$, $R_\text{d}{=}0{.}1$ $\mu$m, $v{=}1$ $\mu$m$/s$, $p_\text{antero}{=}1$. For biologically reasonable transition rates $k_{m\rightarrow w}{=}10/$s and $k_{w\rightarrow m}\in\{10^{-1}/\text{s};10^{0}/\text{s};10^{1}/\text{s}\}$ the reaction mode (left-hand side for reaction by pure encounter and right-hand side for reaction which is only possible in the waiting state) has no major impact on the search efficiency.
\textbf{a} 
The optimal width of the actin cortex is approximately $\delta^\text{opt}{=}0{.}3$ $\mu$m in the case of a uniform distribution of actin filaments (according to Eq. \ref{rotangledistr3D}) in spherical cells.
\textbf{b} 
Influence of cell shape on the search efficiency of transport tasks in spheroidal cells with semi-axes $c > a$ and various eccentricities $\epsilon$.
\textbf{c}
The MFPT for space-dependent transition rates $\tilde{k}_{w\rightarrow m}(r)$ and a total mean number of microtubules in a spherical cell of $N_\text{MT}\in\{10^2;10^3\}$ in comparison to constant waiting times.
\textbf{d}
Impact of gaussian distributed actin orientations with mean $\mu\in\{0;\pi/2;\pi\}$ and standard deviation $\sigma{=}1$ on the MFPT in comparison to a uniformly random actin network for spherical cells. 
}
\label{figure6}
\end{figure*}

\subsection*{Towards a more realistic Cytoskeleton}

\noindent
So far we have studied intracellular search strategies in two-dimensional environments. 
However, a generalization of our model to three-dimensional spherical cells is straightforward.
When updating the particle direction, the new velocity $\boldsymbol{v}^\star(v,\theta_v^\star,\varphi_v^\star)$ is given in a rotated coordinate system where the $z^\star$-axis is pointing in the direction of the current particle position $\boldsymbol{r}(r,\theta_r,\varphi_r)$. While $\varphi_v^\star$ is uniformly distributed in $[0;2\pi)$, the azimuth $\theta_v^\star\in[0;\pi]$ is drawn from the rotation angle distribution 
\begin{equation}
f_\text{3D}(\alpha_\text{rot})=
\begin{cases}
p_\text{a.}\,\delta(\alpha_\text{rot})\!\!\!\!\!\!&+\,(1{-}p_\text{a.})\,\delta(\alpha_\text{rot}{-}\pi), \\
&\text{ for } 0<r<R_\text{m}{-}\delta,\\
1/\pi, & \text{ for } R_\text{m}{-}\delta<r<R_\text{m},
\end{cases}
\label{rotangledistr3D}
\end{equation}
which is the 3D complement of Eq. \ref{rotangledistr}.
In Fig. \ref{figure6} a, c, d we assume a three-dimensional spherical cell of radius $R_\text{m}{=}5$ $\mu$m which is consistent with the dimension of T cells. Since the diameter of an immunological synapse is reported to be of the order of microns \cite{Grakoui1999,Qu2011,Angus2013,Ritter2013}, we choose $\alpha_\text{exit}{=}0{.}2$ which results in an exit zone of size $0{.}2\times 5$ $\mu$m ${=}1$ $\mu$m. We further take into account the typical size of vesicles by $R_\text{d}{=}0{.}1$ $\mu$m \cite{Qu2011} and the conventional motor speed by $v{=}1$ $\mu$m$/$s \cite{TheCell}. Since cargo typically detaches at network intersections \cite{Balint2013}, a transition rate of $k_{m\rightarrow w}{=}v/\ell{=}10/$s results in a biologically reasonable mesh size of $\ell{=}100$ nm \cite{Eghiaian2015,Salbreux2012}. We consider $k_{w\rightarrow m}\in\{10^{-1}/\text{s};10^{0}/\text{s};10^{1}/\text{s}\}$, because the mean waiting time of cargo at network intersections is reported to be of the order of seconds \cite{Balint2013,Slepchenko2007,Zaliapin2005}, and choose $p_\text{antero}{=}1$ which is optimal for the reaction-escape problem as found in Fig. \ref{figure5} and reflects a low activity level of dynein motors.\\

\noindent
Under these circumstances, Fig. \ref{figure6} a reveals an optimal width of the actin cortex of approximately $\delta^\text{opt}{=}0{.}3$ $\mu$m. This is in good agreement to the typical width of an actin cortex \cite{Eghiaian2015,Salbreux2012,Clark2013}. Remarkably, for biologically reasonable parameters the reaction mode has no significant effect on the search efficiency, as already expected by Fig. \ref{figure2} e (here, the dimensionless transition rate is $k_{m\rightarrow w}\mapsto R_\text{m}k_{m\rightarrow w}/v{=}50$). Moreover, the divergence of the MFPT in the limit of $\delta\to 0$ outlines the role of the actin cortex between functional gateway (optimized $\delta$) and transport barrier ($\delta\to 0$) \cite{Papadopulos2013}.\\

\noindent
Still, Eq. \ref{rotangledistr3D} reflects a simplified view of the real cytoskeleton. Next, we take three steps towards a more realistic cytoskeletal organization.\\
Until now we investigated non-polarised spherical cells. However most cells show some sort of polarisation \cite{TheCell}. In order to check for the influence of polarisation on the search efficiency of intracellular transport we investigate ellipsoidal cells. The cell shape is modeled by a prolate spheriod with $z$ as the symmetry axis and semi-axes $c$, $a$ ($c>a$). In Fig. \ref{figure6} b we fix the volume of the spheroid $V{=}4 \pi (5\,\mu\text{m})^3/3$ in order to exclude volume effects and vary the eccentricity $\epsilon{=}\sqrt{1-a^2/c^2}\in[0;1[$, where $\epsilon{=}0$ corresponds to a spherical cell. For the narrow escape problem the exit zone is located at the north pole of the spheroid. Note that $\alpha_\text{exit}$ is adapted to the eccentricity $\epsilon$ to guarantee a fixed target size of $1\,\mu$m. Under biologically reasonable conditions Fig. \ref{figure6} b indicates that the gain in search efficiency by a thin actin cortex is conserved for ellipsoidal cells.\\

\noindent
Next we consider that the concentration of microtubules is not constant but depends on the radial position $r$ within the cell. While the microtubule network is more dense close to the MTOC, it gets more and more dilute in the cell periphery. The probability to find a microtubule decreases with increasing $r$ and can be estimated by $\rho_\text{2D}^{~}(r)=N_\text{MT}^{~}d_\text{MT}^{~}/(2\pi r)$ in 2D and $\rho_\text{3D}^{~}(r)=N_\text{MT}^{~}d_\text{MT}^2/(16 r^2)$ in 3D, where $N_\text{MT}{=}10^2{-}10^3$ specifies the mean number of microtubules in the cell \cite{Cohen1970,Schulze1987} and $d_\text{MT}{=}25$nm is the diameter of a single microtubule \cite{TheCell}.
This can be considered in our model by introducing position-dependent transition rates $\tilde{k}_{w\rightarrow m}(r){=}\rho (r)\!\cdot \!k_{w\rightarrow m}$ in the cell interior. Consequently, the greater the distance from the MTOC and thus the lower the microtubule density, the longer the particle remains in the waiting state.
Fig. \ref{figure6} c shows the MFPT for space-dependent transition rates $\tilde{k}_{w\rightarrow m}(r)$ in comparison to the results for constant transition rate $k_{w\rightarrow m}$. As expected, the impact of the variable microtubule density on the search efficiency vanishes with increasing width of the actin cortex $\delta$, since in the periphery the rate $k_{w\rightarrow m}$ is constant. Furthermore, the effect is stronger for low values of $N_\text{MT}$ due to the resulting greater mean waiting time per waiting period. Overall, the gain in efficiency by a thin actin cortex is conserved for the reaction-escape problem.\\

\noindent
Eventhough the orientation of actin filaments in the cortex typically is random, the exact distribution of actin polarities is elusive. There are several mechanisms which influence the directionality of actin networks. For instance, actin filaments align to microtubules \cite{Lopez2014} or form branches at distinct angles by the protein complex Arp2/3 \cite{Mullins1998,Risca2012}. In Fig. \ref{figure6} d, we analyze non-uniform, cut-off-gaussian rotation angle distributions in the cell cortex
\begin{equation}
\tilde{f}_\text{3D}(\alpha_\text{rot})=
\begin{cases}
p_\text{a.}\,\delta(\alpha_\text{rot})+(1{-}&\!\!\!\!\!\!p_\text{a.})\,\delta(\alpha_\text{rot}{-}\pi), \\
&\text{ for } 0<r<R_\text{m}{-}\delta,\\
\frac{\mathcal{N}}{\sqrt{2\pi}}e^{{-}\frac{1}{2}(\alpha_\text{rot}{-}\mu)^2}, & \text{ for } R_\text{m}{-}\delta<r<R_\text{m},
\end{cases}
\label{rotangledistr3D_b}
\end{equation}
with mean $\mu\!>\!0$ and normalization $\mathcal{N}$ so that $\alpha_\text{rot} \in [0;\pi]$. A mean value $\mu\!\in\![0;\pi/2)$ ($\mu\!\in\!(\pi/2;\pi]$) leads to outwardly (inwardly) polarized actin filaments, whereas $\mu{=}\pi/2$ corresponds to a predominantly lateral orientation.
Fig. \ref{figure6} d displays the MFPT in dependence of the cortex width $\delta$ for diverse polarities of the actin cortex defined by the mean values $\mu\in\{0;\pi/2;\pi\}$.
The behavior of the MFPT for lateral actin orientations ($\mu{=}\pi/2$) is similar to the one of uniformly random actin networks, because of the equal probability for outward- and inward-directed cargo transport. An inhomogeneous cytoskeletal structure with a thin actin cortex generally supports the search efficiency.
To the contrary, for outward-pointing actin filaments ($\mu\!<\!\pi/2$) the particle is predominantly moving in close proximity of the cell boundary due to the actin network polarity. For that reason the cortex width $\delta$ largely does not influence the MFPT and the homogeneous limit $\delta{=}1$ is most efficient.
Inward-directed actin polarities ($\mu\!>\!\pi/2$) drive the cargo towards the cell center. While this is advantageous for the pure reaction problem (the effect is comparable to $p_\text{antero}{=}0$ in Fig. \ref{figure3}), it is highly disadvantageous for the escape and consequently also for the reaction-escape problem. There, an efficient search strategy strongly depends on the formation of a thin actin cortex which forces the particle closer to the membrane, as evident from the pronounced minimum of the MFPT.
In general, a larger standard deviation of the gaussian in Eq. \ref{rotangledistr3D_b} randomizes the actin network. Consequently, the behavior of the MFPT converges to the uniform case and an inhomogeneous cytoskeleton generally improves the search efficiency again (see also \cite{Hafner2016}).

\section*{Discussion}

\noindent
We analyzed the importance of the spatial organization of the cytoskeleton of living cells for targeted intracellular transport, which occurs when cargo particles have to find reaction partners or specific target areas inside a cell. 
Motor-assisted transport along cytoskeletal filaments is known to enhance reaction kinetics in 
cells which are homogeneously filled with random filaments \cite{Loverdo2008}. But such a condition is only fulfilled in a thin actin cortex underneath the plasma membrane, since the interior of the cell allows merely radial transport along microtubules.\\

\noindent
Remarkably, we find that the confinement of randomly oriented cytoskeletal filaments to a thin actin cortex is not a handicap for the cell, but can substantially increase the efficiency of diverse transport tasks. 
We obtain this result by formulation of a random velocity model with intermittent arrest states which takes into account the spatially inhomogeneous structure of the cytoskeleton. 
Our model allows the systematic study of the impact of the cytoskeleton organization, the motor behavior and the target detection mode on the search efficiency of diverse transport tasks. 
Here, we analyze three paradigmatic intracellular search problems: 
the narrow escape problem, the reaction problem, and the reaction-escape problem.\\

\noindent
We find that the best strategy for the escape problem is to transport the cargo towards the periphery by a high probability for anterograde transport and keep it in close proximity of the membrane by providing a thin actin cortex.
In the case of the reaction problem efficient detection is based on bringing the reaction partners with identical motility patterns close together either at the center of the cell ($p_\text{antero}{=}0$, $\delta{=}0$) or in a thin cortex ($p_\text{antero}{=}1$, $\delta^\text{opt}$). However, this strategy fails for nonidentical particles with different motor activity levels defined by $p_\text{antero}$. Then, a homogeneous, random cytoskeleton ($\delta{=}1$) is advantageous. While for the reaction problem $p_\text{antero}{=}0$ is most efficient, it is highly deficient for the escape problem. Hence, both strategies have to be combined effectively for the reaction-escape problem. In the case that the motility pattern of the product particle is not different from the one of searcher and target, a high probability for anterograde transport and establishing a thin actin cortex is favorable. Table \ref{table1} recapitulates the best search strategies for the three transport tasks under investigation.\\

\noindent
Our results indicate that cells with a centrosome are able to realize efficient intracellular search strategies by intermittent transport on a cytoskeleton with specific spatial structure (see also \cite{Schwarz2016,Schwarz2016b} for similar findings in a model with intermittent diffusion).
In comparison to the homogeneous pendant, an inhomogeneous cytoskeleton organization which displays only a thin actin cortex generally leads to a considerable gain in search efficiency for diverse intracellular transport tasks.

\begin{table*}[t]
\centering
\begin{tabular}{l m{2cm} m{2cm} m{2cm} m{2cm}}
~ & \multicolumn{2}{c}{\textbf{homogeneous cytoskeleton~~~~~~~}} & \multicolumn{2}{c}{\textbf{inhomogeneous cytoskeleton}} \\
~ & $~~~~~~k_{m\to w}^\text{opt}$ & $~~~~~~k_{w\to m}^\text{opt}$ & $~~~~~~p_\text{antero}^\text{opt}$ & $~~~~~~\delta^\text{opt}$ \\
\hline
\textbf{narrow escape} & $~~~~~~0$ & $~~~~~~\infty$ & $~~~~~~1$ & $~~~~~~\in\,\,]0;1[$ \\
\hline
\textbf{reaction of identical particles} & ~ & ~ & ~ & ~ \\ 
detection: encounter & $~~~~~~0$ & $~~~~~~\infty$ & $~~~~~~0$ & $~~~~~~0$ \\
detection: waiting & $~~~~~~{\approx}13$ & $~~~~~~{\approx}34$ & $~~~~~~0$ & $~~~~~~0$ \\
\hline
\textbf{reaction with immotile target} & ~ & ~ & ~ & ~ \\ 
detection: encounter & $~~~~~~0$ & $~~~~~~\infty$ & $~~~~~~0{.}5$ & $~~~~~~1$ \\
detection: waiting & $~~~~~~{\approx}7$ & $~~~~~~\infty$ & $~~~~~~0{.}5$ & $~~~~~~1$ \\
\hline
\textbf{reaction-escape of identical particles~~} & ~ & ~ & ~ & ~ \\ 
detection: encounter & $~~~~~~0$ & $~~~~~~\infty$ & $~~~~~~1$ & $~~~~~~\in\,\,]0;1[$ \\
detection: waiting & $~~~~~~{\approx}5$ & $~~~~~~{\approx}22$ & $~~~~~~1$ & $~~~~~~\in\,\,]0;1[$ \\
\hline
\end{tabular}
\caption{
Recapitulation of the optimal search strategies for narrow escapes defined by $\alpha_\text{exit}{=}0{.}1$ and reaction partners of size $R_\text{d}{=}0{.}1$. The optimal transition rates $k_{m\to w}^\text{opt}$ and $k_{w\to m}^\text{opt}$ of a homogeneous cytoskeleton structure are applied for the inhomogeneous case.
}
\label{table1}
\end{table*}

\section*{Author Contributions}

\noindent
H.R. designed the research. A.E.H. performed calculations, prepared figures and analyzed the data. H.R. and A.E.H. wrote the manuscript.

\section*{Acknowledgements}

\noindent
The authors wish to thank Gleb Oshanin for helpful discussions. This work was financially supported by the German Research Foundation (DFG) within the Collaborative Research Center SFB 1027.

\end{document}